# Mining Insights on Metal-Organic Framework Synthesis from Scientific Literature Texts

Hyunsoo Park[†,⊥], Yeonghun Kang[†,⊥], Wonyoung Choe[‡], and Jihan Kim[†,*]

† Department of Chemical and Biomolecular Engineering, Korea Advanced Institute of Science and Technology (KAIST), 291, Daehak-ro, Yuseong-gu, Daejeon 34141, Republic of Korea

‡ Department of Chemistry, Ulsan National Institute of Science and Technology (UNIST), 50, UNIST-gil, Eonyang-eup, Ulju-gun, Ulsan, 44919, Republic of Korea

⊥These authors contributed equally to this work

# Abstract

Identifying optimal synthesis conditions for metal-organic frameworks (MOFs) is a major challenge that can serve as a bottleneck for new materials discovery and development. Trial-and-error approach that relies on a chemist's intuition and knowledge has limitations in efficiency due to the large MOF synthesis space. To this end, 47,187 number of MOF were data mined using our in-house developed code to extract their synthesis information in 28,565 MOF papers. The joint machine learning/rule-based algorithm yields an average F1 score of 90.3 % across different synthesis parameters (i.e. metal precursors, organic precursors, solvents, temperature, time, composition). From this data set, a PU learning algorithm was developed to predict synthesis of a given MOF material using synthesis conditions as inputs, and this algorithm successfully predicted successful synthesis in 83.1 % of the synthesized data in the test set. Finally, our model correctly predicted three amorphous MOFs (with their representative experimental synthesis condition) as having low synthesizability scores while the counterpart crystalline MOFs showed high synthesizability scores. Our results show that big data extracted from the texts of MOF papers can be used to rationally predict synthesis conditions for these materials, which can accelerate the speed in which new MOFs are synthesized.

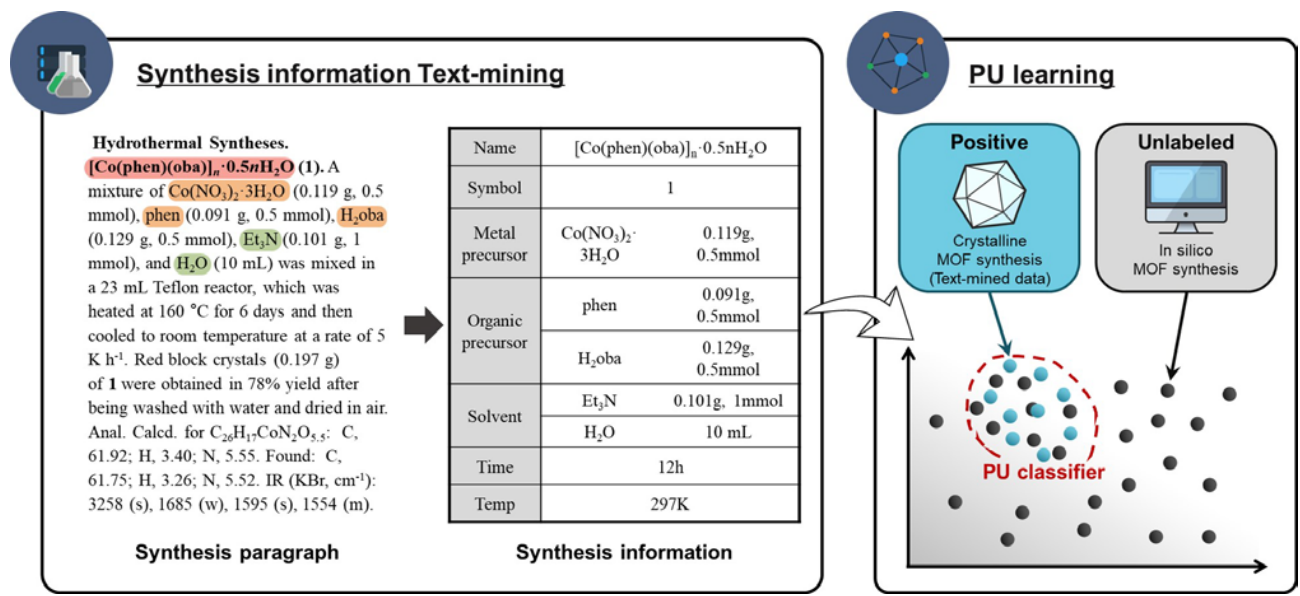

TOC figure

# INTRODUCTION

Identifying ideal synthesis conditions for nanomaterials is one of the most important endeavours when it comes to materials discovery and development. Given the large number of tunable synthesis parameters (e.g. temperature, composition, reaction time, solvent selections, and pH), the time spent in selecting the proper synthesis conditions often serves as the bottleneck of the research. For the most part, heuristics and knowledge-based intuitions are primarily used by the experimentalists to find the optimal synthesis conditions. However, this type of an approach can be inefficient and rely upon the expertise of the researcher, which can lead to large variance in the quality of the synthesized materials.

To expedite this process, researchers have started to use various machine learning and AI algorithm to predict materials synthesis using a large amount of available synthesis data.[1-10] In particular, Raccuglia et al.[1] successfully demonstrated that the failed experiments collected from archived laboratory notebooks can be used to predict synthesis of materials (i.e. the crystallization of templated vanadium selenites) using machine learning. Also, Ahneman et al.[2] developed a random forest model to predict high-yield reaction conditions in C-N cross coupling by training data obtained via high-throughput experiments. And finally, Walker et al.[3] developed a machine learning model to predict experimental synthesis conditions of organic reactions. The model was trained using Reaxys, which is a large organic synthesis database for over 45 million reactions.

On the other hand, texts found in the scientific literature is one potential source of data that has yet to be fully exploited to predict the synthesis conditions. Given that millions of nanomaterials and nanotechnology-related papers exist in the publication archive, one can in principle use this aggregated information to potentially optimize the synthesis procedure. Unfortunately, this data is not provided in an organized manner and as such, various mining tools and algorithm need to be developed to collect this information. In the seminal paper that connected text mining with materials development, Tshitoyan et al.[11] demonstrated that the word embedding encoded from scientific papers enabled learning chemical knowledge. Also, Kim et al.[12] collected the synthesis parameters of oxide materials compiled across 30 systems from scientific papers. Finally, Kononova et al.[13] collected and provided the synthesis conditions for about 20,000 reactions from over 50,000 solid-state synthesis paragraphs. Moreover, there have been numerous other papers that used various text mining tools to collect useful information from materials related papers.[14-17]

However, to the best of our knowledge, there hasn't been any work that used published experimental text data to predict synthesizability for materials given synthesis conditions as inputs. To this end, we have collected over 28,565 papers on a materials class called metal-organic frameworks (MOFs) registered in the Cambridge Structure Database metal-organic framework subset (CSD MOF subset) with the hypothesis that large amounts of synthesis data can reveal useful information about the synthesis procedure. In general, MOFs are renowned for their high surface area and tunable properties, and due to facile synthesis procedures, there are over 100,000 MOFs that have been synthesized thus far.[18] Moreover, various parameters such as metal and organic precursor types, composition, temperature, time, and solvents play a key role in the MOF synthesis. As such, we developed a text mining algorithm that accurately collects data related to synthesis for 47,187 MOF synthesis conditions. Using this information, an artificial neural network (ANN) model was trained using positive-unlabeled learning (PU learning) to predict synthesizability for a given input synthesis conditions. Our ANN model shows high predictability of synthesis and moreover can differentiate between amorphous and crystalline form of the same MOF, with the latter possessing a higher synthesis score. This research is a first step towards using massive amounts of MOF synthesis data to facilitate ideal synthesis conditions, which can accelerate materials discovery and development within this field.

# RESULT

## 1. Text-mining

### 1.1 Text-mining Algorithm

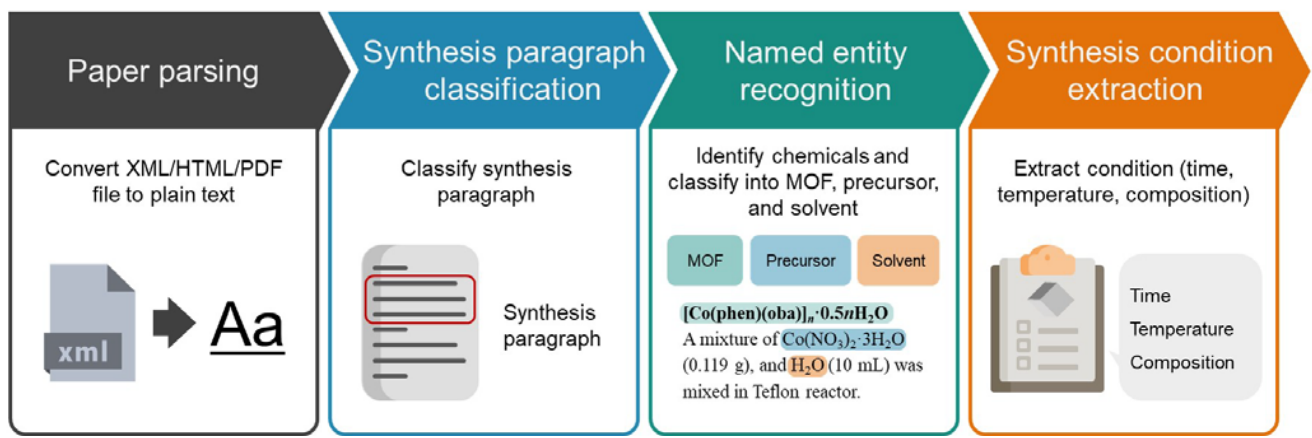

Figure 1. A pipeline of text-mining algorithm that extracts synthesis conditions from MOF papers. It consists of (1) paper parsing (2) synthesis paragraph classification (3) named entity recognition and (4) synthesis condition extraction.

Figure 1 shows a pipeline of automatically extracting synthesis conditions using our in-house developed text-mining code, and is divided into four parts: (1) paper parsing (2) synthesis paragraph classification (3) named entity recognition and (4) synthesis condition extraction. For (1), MOF papers were obtained in HTML/XML/PDF formats with the permissions from journals. Afterwards, these papers were converted to plain text format using the parsers that extract paragraphs and headings in these files. For (2), machine learning was used to identify paragraphs that contain parts relevant to experimental synthesis (details of the algorithm can be found in the Methods section). The results of different machine learning methods (i.e. logistic regression, soft vector machine, and random forest) are summarized in Table S1, and while all of the methods yielded a high precision of over 98 %, the recall scores varied significantly (90, 85, and 67 % for logistic regression, soft vector machine, and random forest, respectively). Therefore, we opted to choose the logistic regression method to identify the synthesis paragraphs. Next, it is imperative to extract useful data related to synthesis conditions from these paragraphs. As such, for (3), the chemicals in MOF papers were collected with the named entity recognition (NER). NER is an application of natural language processing that extracts entities and properly puts them into different categories such as locations and names. Recently, NER has been used in chemistry and materials sciences to facilitate extracting chemicals and material properties from texts.[13,19,20] In this work, NER was trained to identify and classify the chemicals into MOF names (e.g. IRMOF-1, HKUST-1, and MOF-74), precursor (e.g. $Co(NO_3)_2 \cdot 3H_2O$, BDC, and trimesic acid), and solvents (e.g. water, DMF, and ethanol). The results are summarized in Table S2 with the details of the NER model explained in the Methods section. Finally, for (4), the synthesis conditions (i.e. temperature, time, and composition) were extracted using regular expressions and mapped to the chemicals identified from (3). The details of the rule-based algorithm are explained in the Methods section.

## 1.2. Results of Text-mining MOF Papers

Our text-mining code was developed according to the pipeline in the previous section and it was tested using a test set created from manually extracting the synthesis conditions from 100 randomly selected MOF papers. The precision, recall, and F1 scores for MOF name, metal precursor, organic precursor, solvent, composition, synthesis temperature, and synthesis time are shown in Figure 2(a). It can be seen that the F1 scores for all of the parameters are higher than 80 %, demonstrating that our algorithm performs well in collecting accurate synthesis data. It should be noted that the precision for the extracted names of metal precursors and organic precursor using NER have a relatively low scores of 81 and 78 %, respectively. Given that the results of the NER model shows a high precision score of 93 % when it comes to identifying the names of the precursors as shown in Table S2, their relatively low precision scores are caused by the problems associated with accurately distinguishing between metal and organic precursors.

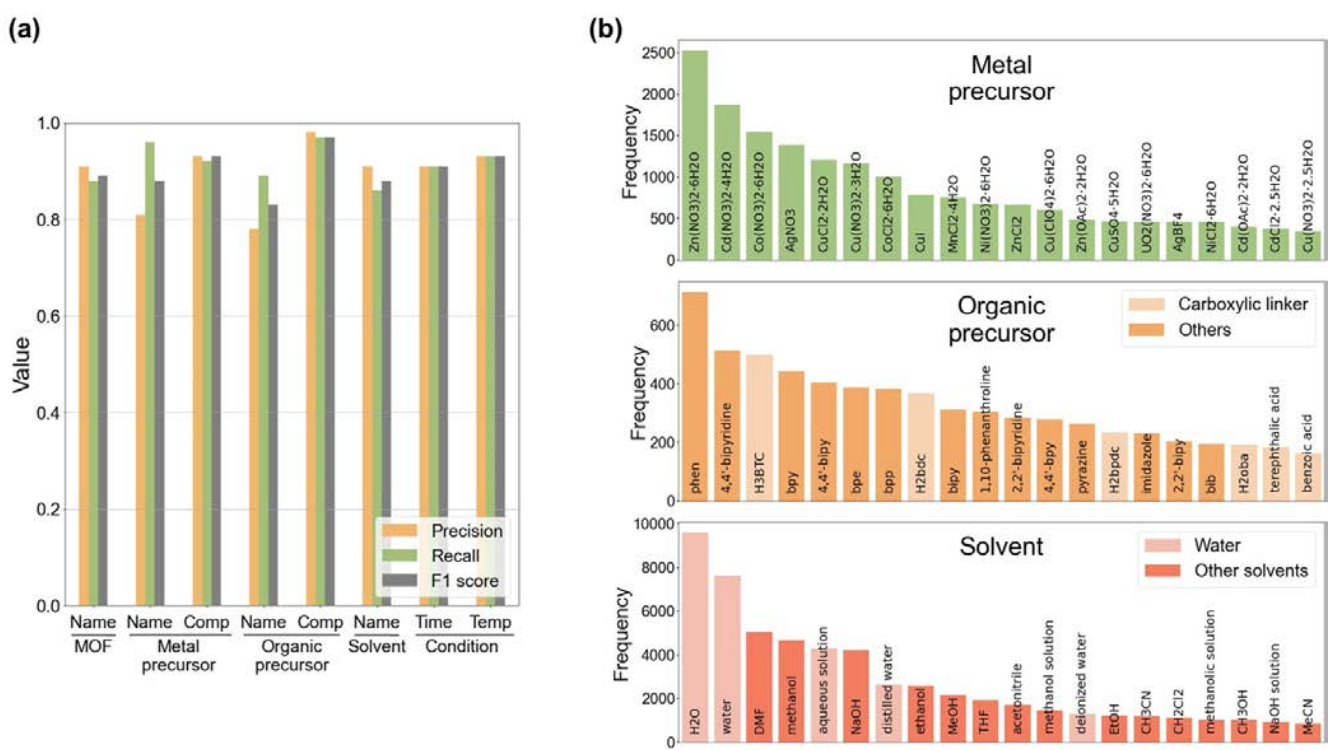

Figure 2. (a) The precision, recall, and F1 scores for MOF name, metal and organic precursor, solvent, composition, time, and temperature. (b) The 20 most frequently occurring metal precursors, organic precursors and solvents extracted from the MOF papers.

Using the text-mining code, synthesis conditions were extracted for 47,187 MOFs from 28,565 papers. Overall, 63,224 metal precursors, 71,042 organic precursors and 93,977 solvents were obtained in the process. Amongst these, the top 20 most frequently cited chemicals are shown in Figure 2(b). It is worth mentioning that all of the MOF papers used in our analysis comes from the CSD MOF subset database that defines MOF based on the building blocks (i.e. metal-containing units, or secondary building units, and the organic linkers) that all MOF scientists agree on.[18] Figure 2(b) shows that the 20 most frequently extracted metal precursors come from following atom types: Cu, Zn, Cd, Co, Ag, and Mn. And this tendency correlates well upon comparisons with the number of structures containing these metal types among the registered structures in the CSD MOF subset (CSD version 5.42) as summarized in Figure

S5(a). For organic precursors, user-defined abbreviations (e.g. L, $L_2$, $L_3$, HL, $H_2L$) can be commonly found within the papers (account for 12.6 % of the overall extracted organic precursors) and to facilitate matter, these were removed from the analysis. The most frequently extracted organic precursors are bipyridine compounds such as phenanthroline (e.g. phen, 1,10- phenanthroline), 4,4'-bipyridine(e.g. 4,4'-bpy, 4,4'-bipy), and 2,2'-bypridine (e.g. bpy, bipy, 2,2'-bipy). This is in contrast to the more commonly known organic precursors like BDC (e.g. H2bdc, terephthalic acid), BTC (e.g. H3btc). The reason is that the CSD MOF database contains many of the old MOFs and thus include many of the organic precursors that are not readily used currently. For solvents, it is interesting to note that water (also labeled as $H_2O$, water, aqueous solution, distilled water, deionized water, etc) is shown to be the most commonly used solvent, and this can be attributed to the fact that most MOFs are synthesized via solvothermal (or hydrothermal) methods. Also, water is used to dilute other solvents like DMF, DMA, and DMSO and to dissolve precursors. Further analysis on the data can be found in Supplementary Note S1-S3.

# 2. PU learning

## 2.1. Results of PU learning

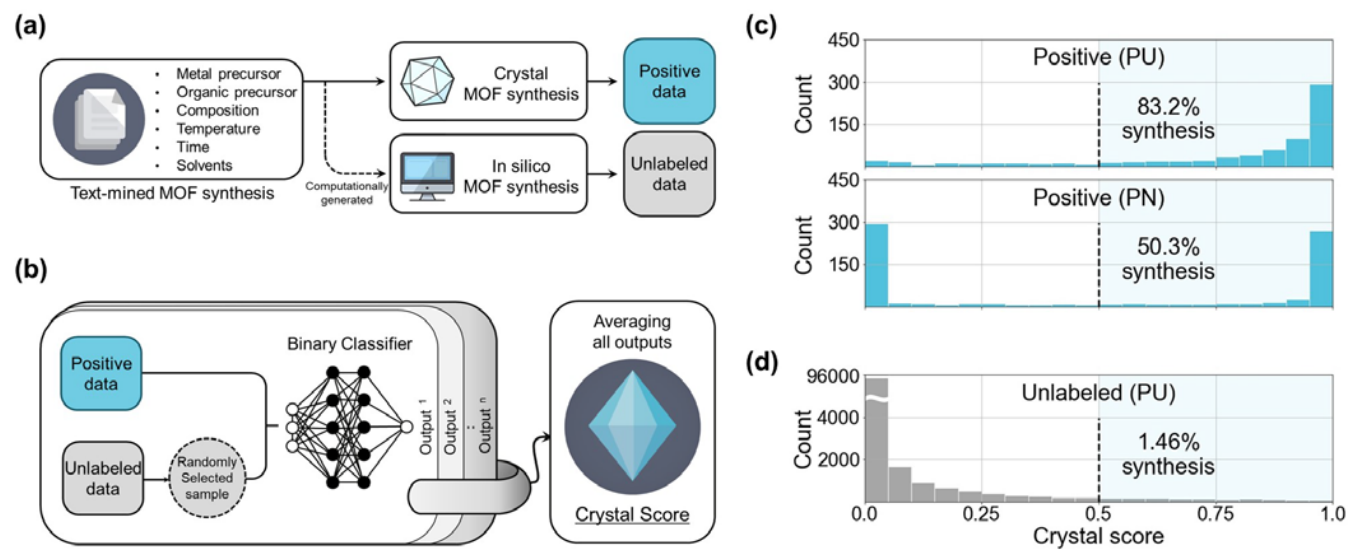

Figure 3. (a) A schematic that differentiates positive (P) and unlabeled (U) data. (b) A schematic outlining our PU learning algorithm. (c) Histograms of prediction score (crystal scores) for positive data in the test set for PU learning and PN learning. (d) Histogram of prediction score (crystal scores) for unlabeled data in the test set with PU learning.

The large amounts of the synthesis conditions obtained using our text-mining code can potentially be used to capture intuitions regarding the synthesizability of the MOFs. Unfortunately, the synthesis conditions written in papers mainly refer to successful experiments, and as such, conditions of failed experiments are difficult to obtain given that most researchers do not publish this data. Subsequently, a conventional binary classification algorithm that differentiates between a balanced proportion of positive and negative data samples cannot be used due to the asymmetry in the published MOF dataset. To remedy this issue, we opted to use a method called PU learning to predict synthesis from the extracted data. In general, PU learning is commonly used when the proportion of positive to negative data is heavily skewed, and the technique has been applied in many applications such as disease gene identification[21,22], spam detection[23], as well as to predict synthesis information in materials such as MXene, inorganic materials.[24,25] Figure 3(a) shows the schematic of generating both the P (positive) and the U (unlabeled) dataset used in our model. Specifically, the synthesis conditions obtained using our text mining algorithm were labeled as the positive data; and the contrasting in-silico synthesis conditions were generated randomly by uniformly sampling each synthesis parameter from the extracted text-mined data. These in-silico synthesis conditions might turn out to be appropriate conditions to generate highly crystalline MOFs, but in the absence of conclusive verification, they are treated as unlabeled data. Figure 3(b) shows the schematic of our PU learning algorithm that originated from the bagging SVM algorithm proposed by Mordelet et al.[26]. In summary, a binary classifier is trained with the positive data and from the randomly selected unlabeled data at each iteration. Upon conducting a large number of iterations, the final prediction score is obtained by averaging the prediction scores of the trained binary classifiers over total number of iterations. At the end, the final prediction score quantifies the likelihood of a given MOF having high crystallinity for a given input condition (with a score of 1 indicating the synthesis conditions for high crystalline structures and the score of 0 indicating the synthesis conditions for low crystalline structures). In this work, the final prediction score is denoted as "crystal score". As such, high and low crystalline structures were classified according to whether the crystal scores of the PU learning model exceed 0.5 or not. The details behind the regarding the PU learning algorithm are found in the Methods section.

To evaluate the performance of the PU learning model, PN learning (positive-negative learning, as known in supervised learning) was implemented as an alternative method that treats the unlabeled data as negative data. For the test set of 750 positive and 100,000 unlabeled data, the recall scores for PU learning and PN learning are 83.1 % and 50.3 % respectively (recall graphs are shown in Figure S6). The histograms of the crystal scores for the PU learning and the PN learning are shown in Figure 3(c). Given that the PU learning has a relatively high recall value, it can be seen that the PU learning can effectively differentiate between labeled and unlabeled data. Moreover, since PN learning has a relatively low recall score, this also implies that the differentiating between the positive and the unlabeled data set is not a trivial task (or else PN performance would have been high as well). As such, positive data with very low crystal scores (i.e. false negative error) is low for PU learning (5.1 % of positive data have crystal score of less than 0.1) while very high for PN learning (41.1 % of positive data have crystal score of less than 0.1). It is worth noting that 1458 data points have a crystal score of over 0.5 in the 100,000 unlabeled data (only 1.46%) as shown in Figure 3(d). As such, the PU learning algorithm predicts that the unlabeled in-silico synthesis conditions will not lead to single-crystalline materials. And this makes sense given that these in-silico conditions are essentially taken randomly from uniformly sampling of the synthesis parameters and will tend to deviate significantly from the original synthesis condition given the large number of parameters that includes the metal precursor, organic precursor, solvent, composition, and synthesis temperature.

## 2.2. Case Study: Crystalline vs Amorphous MOFs

One issue with the PU learning method outlined above is that it is difficult to properly interpret the low synthesizability scores given to the unlabeled data since we do not have any information verifying this prediction. As such, we looked towards some of the reported amorphous MOFs (aMOFs) as guidance on the reliability of our model. In this context, given that we are concerned with high crystallinity, amorphous MOFs can be interpreted as "failed" synthesis, which should in principle lead to low score if our PU model is reliable. In general, there are two different ways to obtain aMOFs: (1) post-processing (via pressure, temperature, chemical treatment, irradiation) and (2) direct synthesis (via solvothermal[27], supersonic cold spraying[28], and sol-gel approach[29]). There are a total of 90 reported aMOFs according to a recently published review paper for aMOF[30]. Given that the synthesis conditions of the MOFs with a solvothermal method are used when training the PU learning, the 70 aMOFs synthesized with post-processing are discarded. Among the remaining 20 aMOFs obtained from direct synthesis, only 11 aMOFs are synthesized via solvothermal method. Disregarding aMOFs that was not found amongst the dataset of the PU learning, only three aMOFs remained and were used as further analysis in this case study.

To facilitate comparison, the synthesis conditions for both the amorphous and the counterpart crystalline MOFs were obtained as shown in Table 1. The nine crystalline MOFs were selected according to Google scholar ranking when searching the name of the three MOFs (i.e. ZIF-8, ZIF-67, Co-MOF-74). The crystal scores of the PU learning algorithm for the aMOFs and the counterpart crystalline MOFs are summarized in Figure 4(a). The crystal scores obtained from our PU learning algorithm for the three aMOFs are relatively low (e.g. 0.224, 0.467, and 0.027) indicating the trained algorithm correctly predicts these structures to be low crystalline. On the other hand, the crystal scores for all of the nine crystalline MOFs are higher than 0.5 except for c3-ZIF-8. The results indicate that there are some signs indicating that the algorithm can correctly classify suitability of synthesis conditions that lead to crystalline vs amorphous MOFs.

To visualize the change in the crystal scores as a function of synthesis parameters, temperature and M/O ratio were varied and the resulting crystal scores were obtained for different hypothetical conditions (see Figure 4(b)). It can be observed from Figure 4(b) that the synthesis parameters of the three aMOFs are located outside the blue (high crystal) region, which is consistent with the data from Table 1. In contrast, it can be seen that beside c3-ZIF-8, all of the crystalline MOFs were within the blue region indicating that the algorithm can accurately predict crystalline vs amorphous MOFs depending on the varying conditions.

Table 1. The synthesis conditions for the amorphous three MOFs and the nine counterpart crystalline MOFs.

| MOF | | Metal precursor | | Organic precursor | | Temperature (°C) | Time (h) | Solvent | Refs |
| --- | --- | --- | --- | --- | --- | --- | --- | --- | --- |
| | | Name | Composition (mmol) | Name | Composition (mmol) | | | | |
| ZIF-8 | **a-ZIF-8** | Zn(oAC)$_2$ | 0.2 | 2-Methyl imidazole | 0.8 | RT | 0.5 | Water | 31 |
| | c1-ZIF-8 | Zn(NO$_3$)$_2$ | 3.9 | | 276.5 | RT | 1 | Water | 32 |
| | c2-ZIF-8 | Zn(NO$_3$)$_2$ | 2.0 | | 2.0 | 140 | 24 | DMF | 33 |
| | c3-ZIF-8 | Zn(NO$_3$)$_2$ | 2.0 | | 2.0 | RT | 24 | Methanol | 33 |
| ZIF-67 | **a-ZIF-67** | Co(NO$_3$)$_2$ | 0.6 | 2-Methyl imidazole | 0.9 | RT | 5 | Water, Methanol | 34 |
| | c1-ZIF-67 | Co(NO$_3$)$_2$ | 1.6 | | 67.0 | RT | 6 | Water | 35 |
| | c2-ZIF-67 | Co(NO$_3$)$_2$ | 1.9 | | 7.5 | 120 | 1 | Methanol | 36 |
| | c3-ZIF-67 | Co(OAc)$_2$ | 0.5 | | 5.0 | 120 | 24 | Water | 37 |
| Co-MOF-74 | **a-Co-MOF-74** | Co(oAC)$_2$ | 4.0 | 2,5-Dihydroxyterephthalic acid | 1.5 | RT | 2 | Methanol | 38 |
| | c1-Co-MOF-74 | Co(NO$_3$)$_2$ | 3.7 | | 1.1 | 120 | 20 | Water, DMF, Ethanol | 39 |
| | c2-Co-MOF-74 | Co(NO$_3$)$_2$ | 3.1 | | 0.9 | 100 | 24 | Water, DMF, Ethanol | 40 |
| | c3-Co-MOF-74 | Co(NO$_3$)$_2$ | 13.1 | | 3.9 | 100 | 24 | Water, DMF, Ethanol | 41 |

\* oAC : acetate group, RT : room temperature

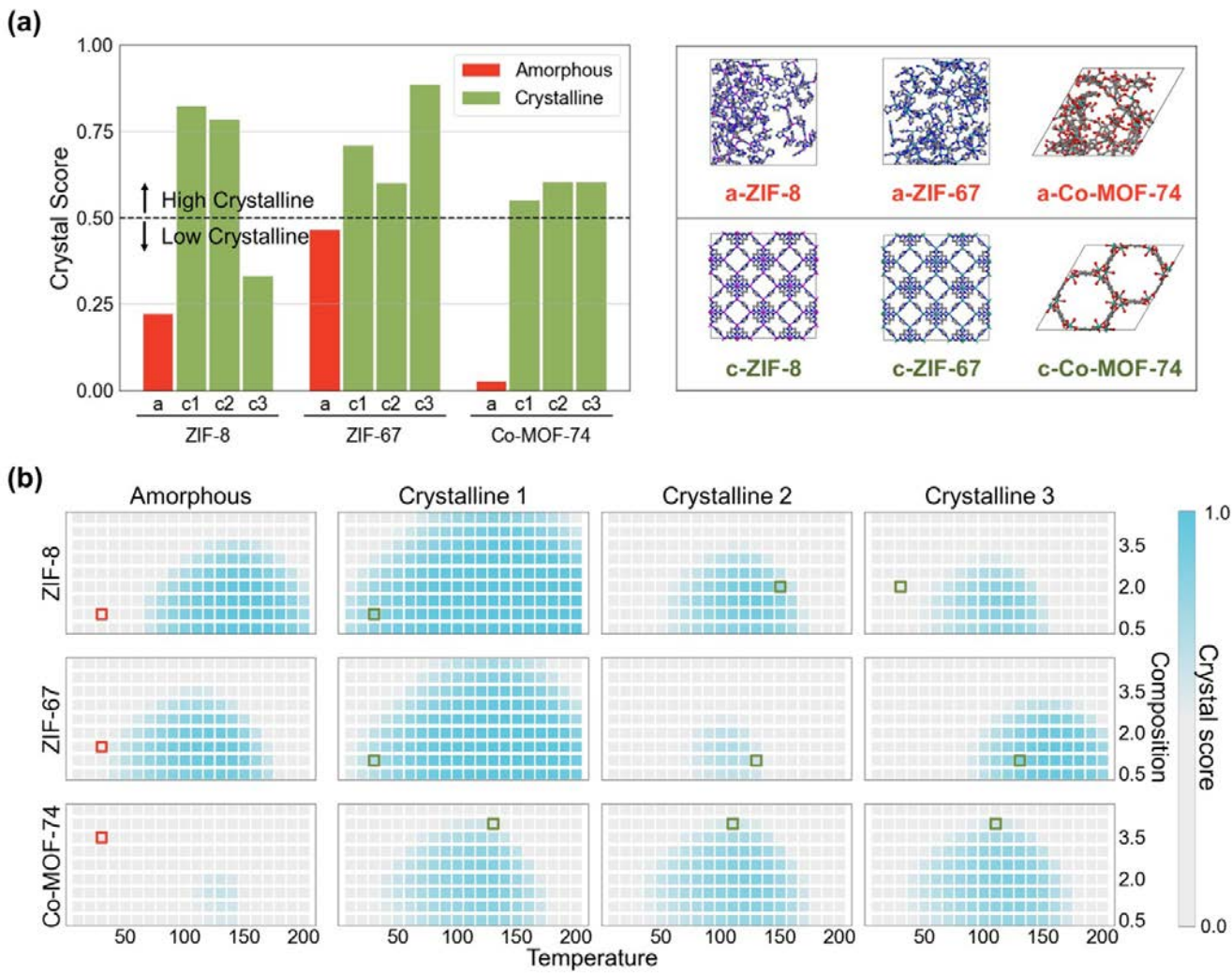

Figure 4. (a) The crystal scores for the amorphous MOFs (red) and the counterpart crystalline MOFs (green) of which synthesis conditions are summarized as Table 1. A value of 0.5 is the boundary dividing high-crystalline and low-crystalline. The figures on the right are drawn to facilitate visualization. (b) The change of the crystal scores with respect to temperature and composition (M/O ratio). Bluer regions indicate high crystal score over 0.5. The red and green marks indicate the synthesis parameters obtained from the published papers for the amorphous and the counterpart MOFs, respectively.

## Conclusions

In this work, an automated text-mining code was developed to extract synthesis conditions from MOF literature. Altogether, total of 47,187 synthesis conditions (i.e. MOF, metal precursor, organic precursor, solvent, temperature, time, composition) were extracted from 28,565 MOF papers. The PU learning model was developed to take the synthesis condition of MOFs as inputs and predict the crystal score as outputs. The model has 83.1 % recall score for positive data in the test set. To verify our model, the synthesis conditions of the reported three amorphous MOFs and the nine counterpart crystalline MOFs are tested and the PU learning model predicted the low crystal scores for all of the aMOFs as well as the high crystal scores for the crystalline MOFs except one. Our findings demonstrate that machine learning method with large amount of data extracted from scientific literatures can help predict synthesis of materials without any prior chemical knowledge. It should be pointed out that there are still some limitations in our current studies: (1) insufficient amounts of data for MOFs with infrequently used metal and/or organic precursors (2) inability to take into account the multi-step synthesis (3) absence of certain parameters such as modulator, pH, stirring time. Nevertheless, we believe our work could be a springboard for predicting synthesis of materials by using big data extracted from scientific literature as guidance to explore chemical space. This will enable accelerated development of MOFs for various new applications.

# METHOD

**Article retrieval**

41,298 digital object identifiers (DOIs) were obtained for MOF papers from the Cambridge Structure Database metal-organic framework subset (CSD MOF subset)[18,42]. Of these, 28,565 papers were downloaded with the permissions from the following five journals: American Chemical Society, Elsevier, Royal Society of Chemical, Springer, and Wiley. Depending on the journal and its supporting format, HTML, XML, and/or PDF files were obtained (Details of formats, journal information are summarized in Table S3).

**Synthesis paragraph classification**

In most of these MOF papers, texts related to synthesis are found in just a few paragraphs. As such, we sought to first identify the paragraphs that contained the synthesis information to narrow our search space. Specifically, texts of 200 MOF papers were divided into paragraphs, and a label of 1 was used to indicate synthesis paragraphs and 0 for other paragraphs. Amongst this set, 180 papers (24,162 paragraphs) and 20 papers (2,930 paragraphs) were used as training and test sets, respectively. Each paragraph was vectorized with the bag-of-words (BOW) representation that keeps track of the multiplicity of the words regardless of grammar and word order using Gensim library[43]. Using the BOW representation as inputs and binary labeling (0 and 1) as outputs, various machine learning methods (i.e., logistic regression, random forest, and soft vector machine methods) were implemented, and F1 scores for each of the methods were obtained using the Sklearn package[44].

**Named entity recognition**

Within the synthesis paragraph, it is important to identify useful chemical information that is critical to the synthesis procedure. For MOF papers, our NER model extracts the chemicals and categorizes them (i.e. MOF, precursor, and solvent) using neural networks. As shown in Figure S7, a 100-dimensional bidirectional-LSTM (Bi-LSTM) which considers forward and backward context was used for the neural network model. Additionally, a conditional random field (CRF) layer is added on top of Bi-LSTM. The CRF layer help to predict each label of sequence data by taking context into account.[45] The embedding layer consists of (1) a 100-dimensional word embedding layer and (2) a 50-dimensional character-leveled word embedding layer. (1) The word embedding was pre-trained with a total of 28,565 MOF papers using word2vec with skip-gram model[46]. (A word embedding analysis is summarized in Supplementary Note S4.) (2) The character-leveled word embedding is an output of a 100-dimensional character-level Bi-LSTM and a 50-dimensional dense layer in the red part of Figure S7. This character-leveled approach can be overcome when unknown words such as new names of MOF, precursor appear.[13]

In order to make a dataset for the NER model, the synthesis paragraphs found from 500 MOF papers were annotated with the BIO (beginning, inside, outside) tags with the categories of chemicals (i.e. MOF, precursor, solvent). The examples and labeling of the BIO tags are summarized in Figure S8. 400/50/50 papers were used as training/test/validation set, respectively. The neural network model was trained with Adam optimizer[47] with 0.001 learning rate and saved weights with the highest accuracy for the validation set during 100 epochs, which was implemented using Tensorflow[48].

**Condition extraction**

Unlike the chemicals, experimental synthesis conditions (e.g. temperature, time, and composition) were obtained using an in-house Python-based rule-based code with the following methods: (1) unit detection method, (2) property classification method, (3) chemical matching method. (1) Unit detection method was developed to identify the units and numerical values of the temperature, time, and composition. (figure S9-(1)). Units can usually be expressed in repeated forms of prefix + main unit + number + suffix. The repeated forms of units were identified using the regular expressions. (2) The extracted units were used with different properties. For example, a unit kelvin (K) is commonly used with various properties like synthesis condition, melting temperature, decomposition temperature, etc. As such, the extracted units were classified into different properties with a rule-based approach which recognizes some keywords in a sentence, and then classifies the most appropriate properties for each unit. (3) The extracted values from (1) need to be mapped with the appropriate chemicals extracted from the NER model (figure S9-(3)). For this, a distance-based method of matching the values with the closest (in terms of distances within the paragraph) chemicals (i.e. MOF, precursor, solvents) was used. As well as synthesis conditions, synthesis operations and methods of MOFs were found using the rule-based code, and the details are explained in Supplementary Note S5.

**Preprocessing data for machine learning**

In this work, a machine learning model was developed to take the synthesis condition of MOFs as inputs and the crystal score as outputs. The synthesis conditions of MOFs need to be preprocessed to use the inputs for machine learning. At first, the names of the extracted precursors from the NER model were classified as metal precursors and organic precursors depending on whether or not metal is included in the name. Then, the names of the precursors need to be vectorized. For example, Benzene-1,4-dicarboxylic acid is one of the most famous organic linkers, but it is written in various names (e.g. BDC, 1,4-Benzenedioic acid, and Terephthalic acid). As such, the names of 754 metal/ 1027 organic precursors with high occurrence were manually replaced with chemical formula and SMILES, respectively. Second, given that more than solvents are mostly used when synthesizing MOFs, the combination of binary inputs was used as inputs with the 5 representative solvents for the synthesis of the MOFs (i.e. water, methanol, ethanol, DMF, DMA). For conditions, the composition was used as the normalized moles of the metal-precursors and the organic precursor to the metal-organic ratio (M/O ratio). Finally, the temperature and M/O ratio values were discretized with intervals of 10 °C and 0.5, respectively. For synthesis time, the time values are often not explicitly written in the papers, especially when synthesized at room temperature. As such, the synthesis time was dismissed to train the machine learning model. It should be noted that there are MOFs commonly synthesized with two organic precursors. As such, the positive data were selected only if they are synthesized with one metal precursor and one or two organic precursors.

**PU learning algorithm**

After preprocessing, the extracted synthesis conditions were used as positive data. The positive data needs to include one more than composition, temperature to be used as input. Also, 155 metal precursors, 423 organic precursors appearing more than respectively 3 times among the vectorized names of the precursors were used. Finally, there is the 3,748 extracted synthesis information of MOFs as the positive data. For the unlabeled data, a total of 1,000,000 data were randomly generated by uniformly sampling each parameter of the positive data. For example, given that the ranges of temperature for the positive data are from 0 to 250 °C, the temperature of the unlabeled data was generated by uniformly sampling between 0 and 250 °C. Finally, 2,998 positive/900,000 unlabeled data and 750 positive/100,000 unlabeled data were used as training and test set, respectively. While machine learning methods such as decision tree, soft vector machine are commonly used as the binary classifiers of the PU learning algorithm (i.e. bagging SVM)[26], the input data in this work is inappropriate to be used with those methods. Because the names of the metal precursor and the organic precursor are sparse and non-continuous data. Instead, the neural network with simple dense layers was used as the binary classifiers. Figure S10 shows the architectures and inputs of the neural networks. The 30 embedding layers were used for each metal/organic precursor. Finally, after they were concatenated, they entered into 32 and 8 dense layers with the elu function[49], a dropout layer with 0.2 rates, and an output layer with sigmoid function. As shown in Figure 3(b), at one iteration, each binary classifier was trained with total positive data and randomly selective data from data unlabeled data of which the number is 10 times the positive data. They were trained with the Adam optimizer[47] during 50 epochs, which was implemented using Tensorflow[48]. The final prediction score was calculated by averaging the scores of all binary classifiers during a total of 30 iterations.

Supplementary information for:

# Mining Insights on Metal-Organic Framework Synthesis from Scientific Literature Texts

Hyunsoo Park[†,⊥], Yeonghun Kang[†,⊥], Wonyoung Choe[‡], and Jihan Kim[†,*]

† Department of Chemical and Biomolecular Engineering, Korea Advanced Institute of Science and Technology (KAIST), 291, Daehak-ro, Yuseong-gu, Daejeon 34141, Republic of Korea

‡ Department of Chemistry, Ulsan National Institute of Science and Technology (UNIST), 50, UNIST-gil, Eonyang-eup, Ulju-gun, Ulsan, 44919, Republic of Korea

⊥These authors contributed equally to this work

# Contents

**Supplementary Note S1.** Frequency distribution of the extracted conditions (composition (M/O ratio), synthesis temperature, synthesis time)

Supplementary Figure S1 shows the frequency distribution of the extracted condition using our text-mining code. The highest frequent M/O ratio value is within a range of 1.0 to 1.25. As well as, it can be seen that the M/O ratio values are widely used in the range near integer multiples near 2.0, 3.0, 4.0, 5.0. For temperature, MOFs are synthesized at a wide range of temperatures between 80 and 190 °C. As well as, they are commonly synthesized at the room temperature range between 20 and 30 °C. Intuitively, it seems reasonable that MOFs are synthesized at various temperatures. The synthesis time value is usually written in the paper on a daily basis such as one day (24 hours), two days (48 hours), and three days (72 hours). The most frequent synthesis time is between 3 and 4 days.

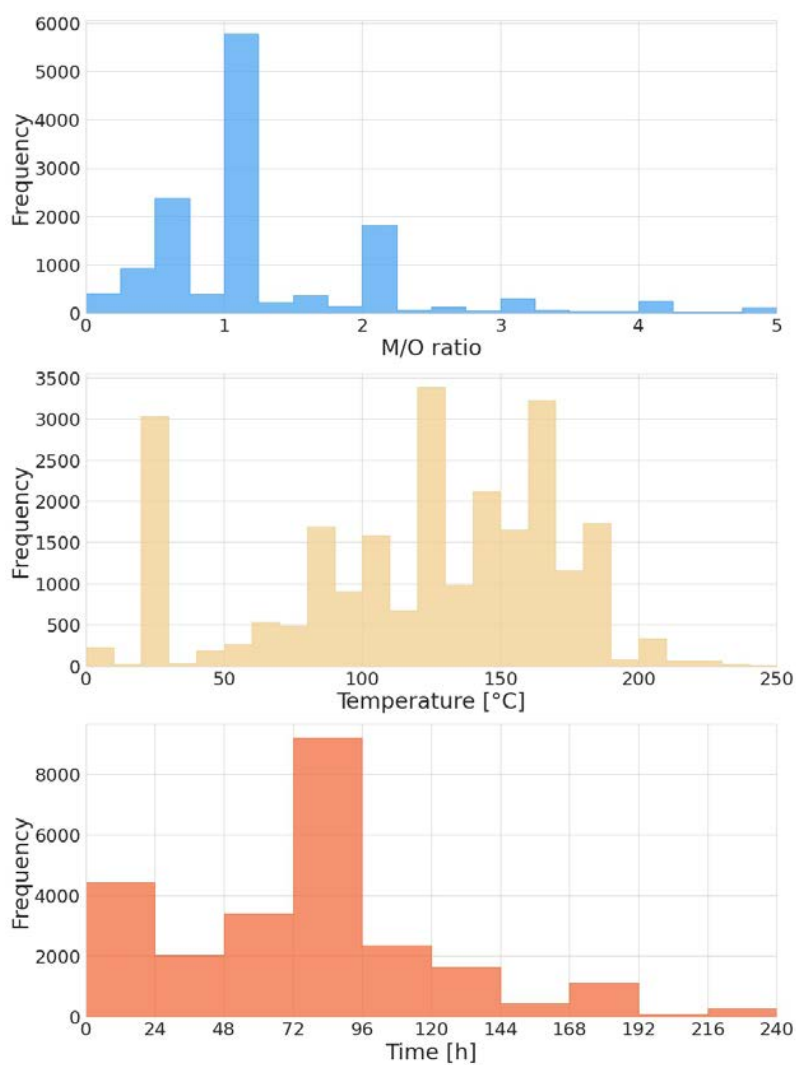

**Supplementary Figure S1**. Frequency distribution of the extracted conditions (composition (i.e. M/O ratio), synthesis temperature, synthesis time)

**Supplementary Note S2.** Synthesis temperature distributions depending metal types of metal precursors.

Supplementary Figure S2 shows density histograms of synthesis temperature concerning the most frequent 9 metal types. The transparent histograms with black lines indicate a density histogram of synthesis temperature for all metals, which is used as background histograms in Supplementary Figure S2. MOFs are mostly synthesized at room temperature (RT, 20~30 °C), 120~130 °C, 160~170 °C. Depending on the metal type, the temperature and median at which MOF is most synthesized are different. The MOFs containing Cu, Ag, Mn, and Fe are synthesized most at RT, whereas Zn, Co, and Ni are synthesized at 120-130 °C, and Cd and Pb are mainly synthesized at 160-170 °C.

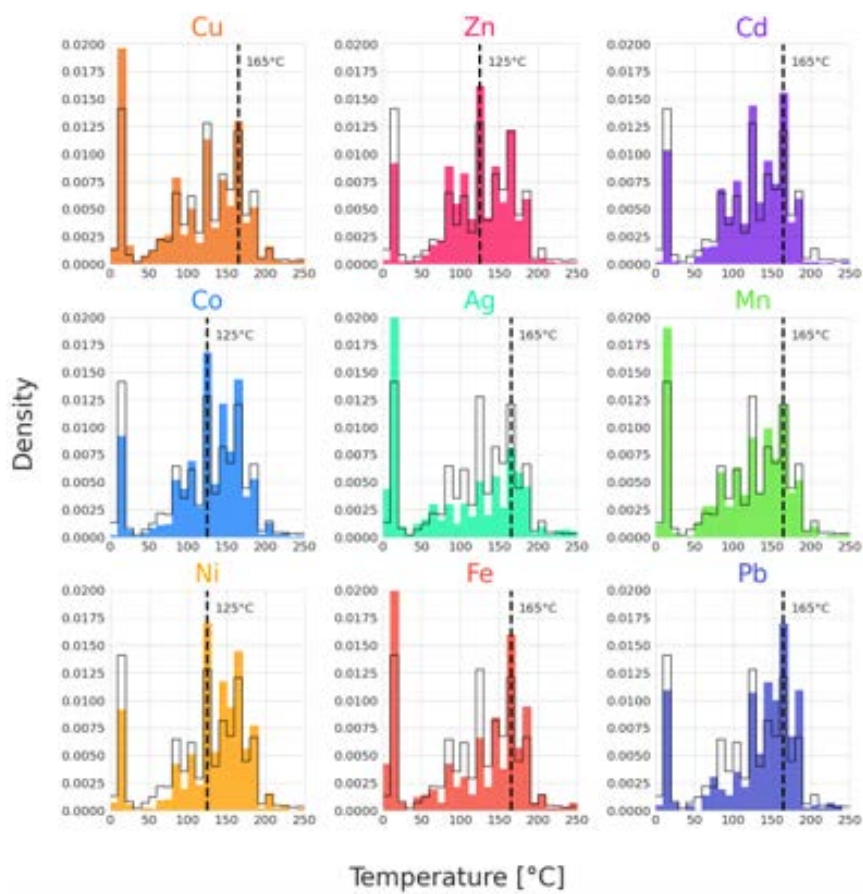

**Supplementary Figure S2**. Synthesis temperature distributions depending metal types of metal precursors (i.e. Cu, Zn, Cd, CO, Ag, Mn, Ni, Fe, Pb).

**Supplementary Note S3.** Synthesis temperature distributions depending solvents.

Supplementary Figure S3 shows violin plots of synthesis temperature for commonly used solvents for MOF synthesis. Supplementary Figure S3(a) is the violin plot when each solvent is used alone (not used with other solvents). Methanol (64.7 °C), ethanol (78.2 °C), acetone (56.1 °C) and acetonitrile (81.3 °C) having lower boiling points than water have median values lower than the boiling point of water (100 °C). Especially, it was observed that methanol and acetone which have the lowest boiling points are mainly synthesized at room temperature. On the other hand, DMF (153.0 °C), DMA (165.1 °C), DMSO (189.0 °C) and DEF (177.5 °C) with higher boiling points than water have median values of 100 °C or higher. They are hardly synthesized at room temperature except for DMSO. Supplementary Figure S3(b) shows a violin plot of synthesis temperature when each solvent is used with water together, which is mostly used with hydrothermal synthesis method. For the hydrothermal synthesis, MOFs are synthesized at high temperatures and pressure. As such, the solvents having low boiling points (methanol, ethanol, acetone, acetonitrile) increase their synthesis temperature. On the other hand, the solvents having high boiling points (DMF, DMA, DMSO, DEF) doesn't change significantly.

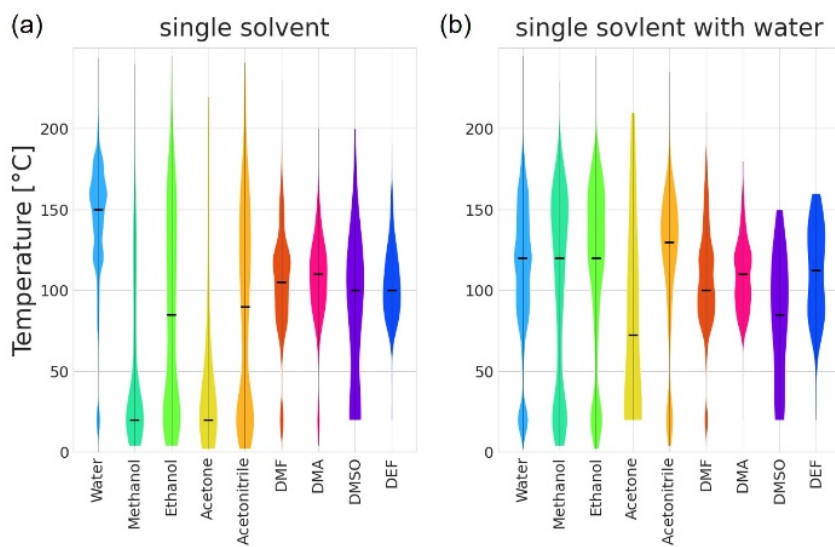

**Supplementary Figure S3.** Violin-histogram of synthesis temperature distributions for (a) single solvent and (b) single solvent with water.

**Supplementary Note S4.** Word embedding analysis for MOF papers.

Word embedding, an effective way to map words into dense and low-dimensional vectors, allows words with similar contexts to have similar vectors.[1,2] In NER, the embedding layer of RNN model (Bi-LSTM CRF) consists of the pre-trained word embedding layer and the character-level word embedding layer. The pre-trained word embedding layer is generated by training tokenized words in the MOF papers. Therefore, the embedding provides an understanding of the relationship between words (i.e. precursor, MOF name, solvent, etc) mostly used in the MOF papers without any prior knowledge.

The analysis of word embedding is generally performed in two methods: (1) semantic similarity and (2) syntactic regularities[3] . (1) semantically similar words lead to be in the nearest (or similar) vector space in the word embedding. Supporting Figure S4(a) shows the semantically similar words in the MOF papers are clustered in a similar position in vector space which is compressed into two-dimension using t-SNE. That is, the categorized words according to names of MOFs, organic precursors, solvents, and units (temperature, time) are grouped in the vector space. (2) The word embedding allows simple semantic operations between words. It enables to semantically operate between MOFs, topology, and metal as shown in Supporting Figure S4(b) drawn by PCA. For example, the following formula is formed between Mg in Mg-MOF-74, and Zn (the commonly used metal of IRMOF-1): [Mg] - [Mg-MOF-74] + [IRMOF-1] = [Zn]. It demonstrates that the word embedding enables to capture of an understanding of relationship and meaning for MOFs without human labeling.

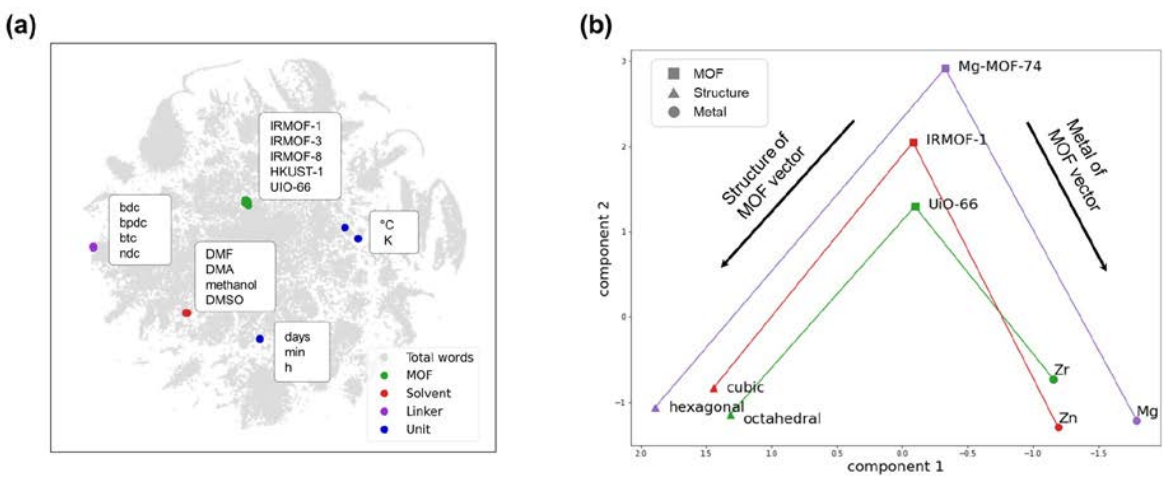

**Supplementary Figure S4.** Word embedding analysis for MOF papers. (a) semantically similar words analysis with t-SNE plot. (b) semantic operations analysis with PCA.

**Supplementary Note S5.** Classification of synthesis methods for MOFs.

   The synthesis method of MOF was divided into six major categories[4], which are classified according to the synthesis method: slow evaporation method, hydro/solvothermal method, microwave-assisted synthesis, sonochemical synthesis, mechanochemical synthesis. At first, 19 operations that are commonly used in MOF synthesis were found by recognizing the keywords using a regular expression and the details are summarized in Table S4. Then, synthesis methods are categorized using a rule-based approach according to the operations. For example, if there is a sonication in methods, it is classified as sonochemical synthesis.

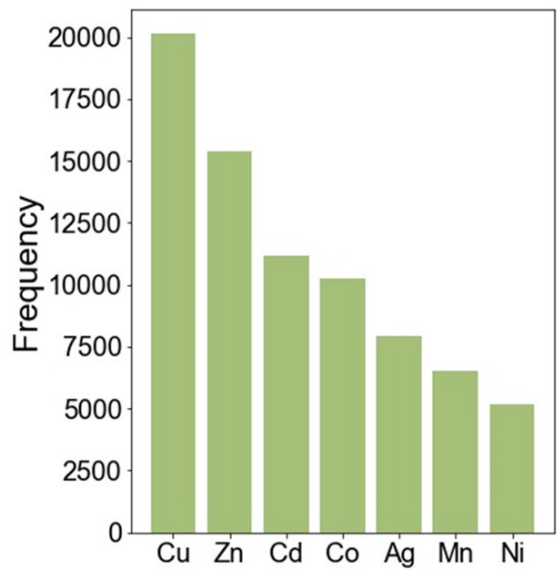
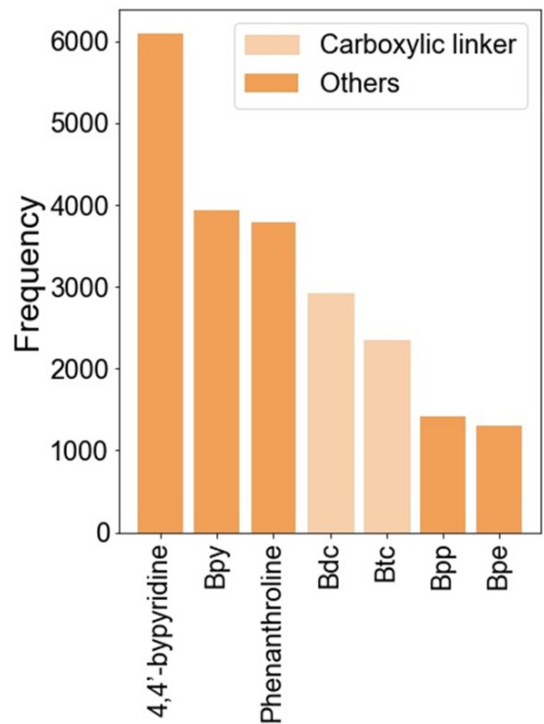

**Supplementary Figure S5**. Histogram of (a) metal frequency and (b) organic linker of structures registered in CSD MOF subset (CSD version 5.42).

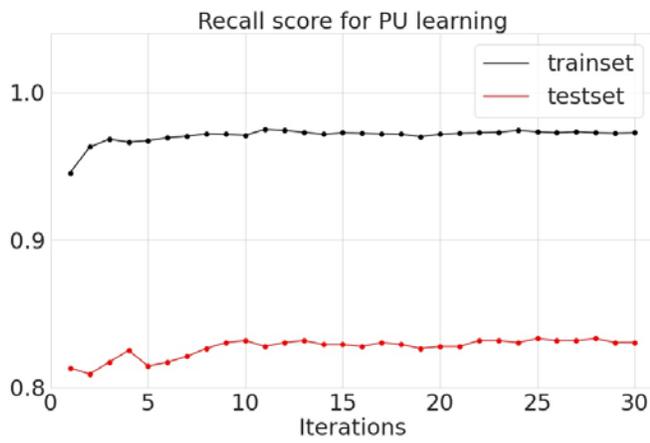 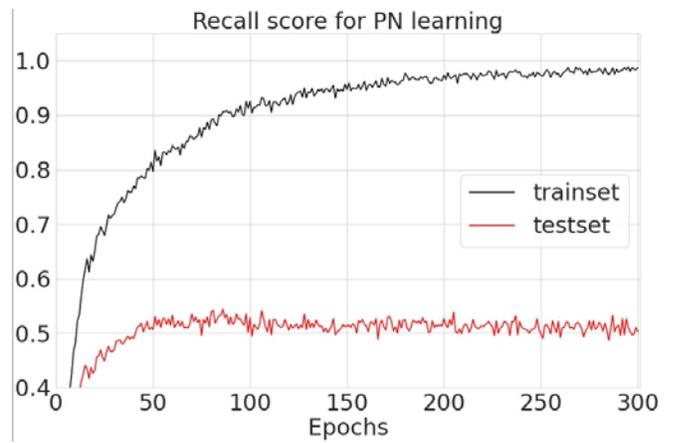

**Supplementary Figure S6**. The curve for recall scores of PU learning (a) and PN learning (b) during training.

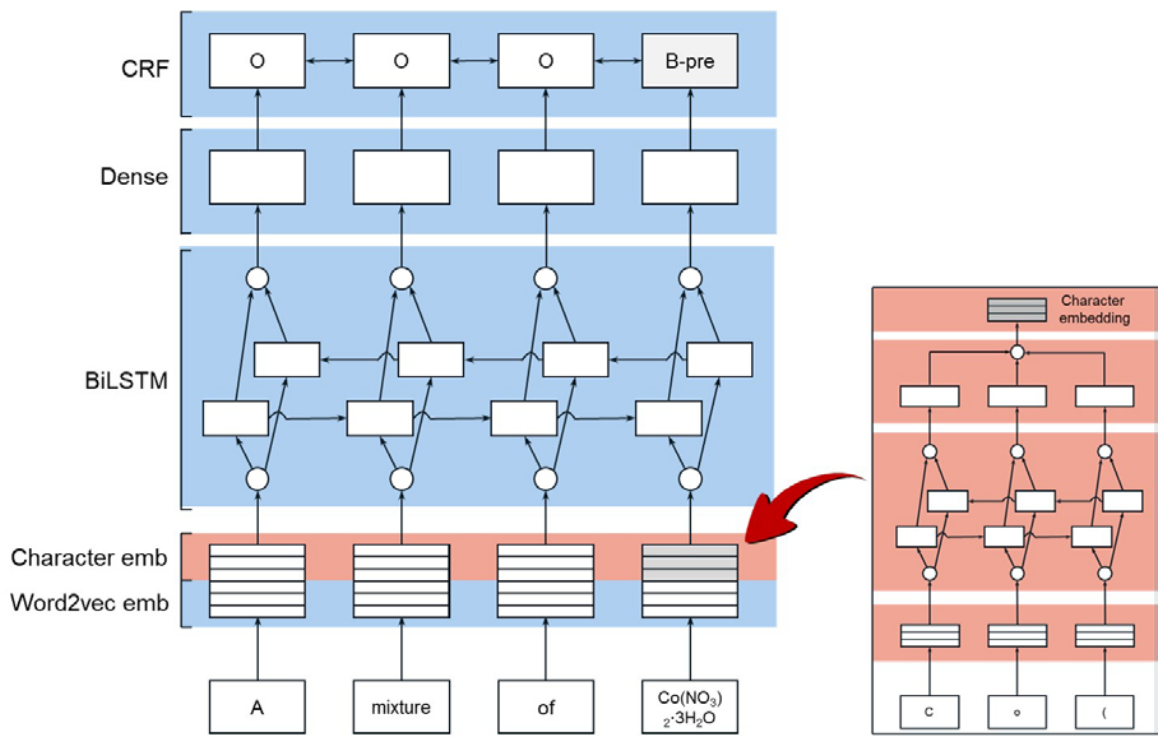

**Supplementary Figure S7**. An RNN architecture of Bi-LSTM CRF for named entity recognition.

Preparation of [CuL1(μ1,3-N3)]n·nClO4. An aqueous solution (2mL) of copper(II) perchlorate hexahydrate (0.1mmol) was added to a methanolic solution (8mL) of L1 (0.1mmol) and stirred at room temperature for 10min.

| Label | Description | Label | Description |
| --- | --- | --- | --- |
| O | Outside | B-precursor | Beginning of the precursor |
| B-MOF | Beginning of the MOF | I-precursor | Inside of the precursor |
| I-MOF | Inside of the MOF | B-solvent | Beginning of the solvent |
|  |  | I-solvent | Inside of the solvent |

**Supplementary Figure S8**. Examples and labeling of BIO tags for MOF, precursor, and solvents.

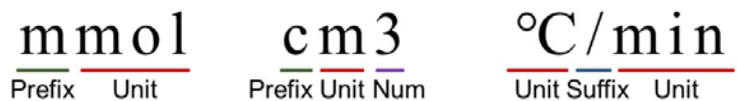

1) **Unit detect method** : Using regular expression, Find unit and value from paragraph Units repeat in (Prefix + Main unit + Suffix + Num) format.

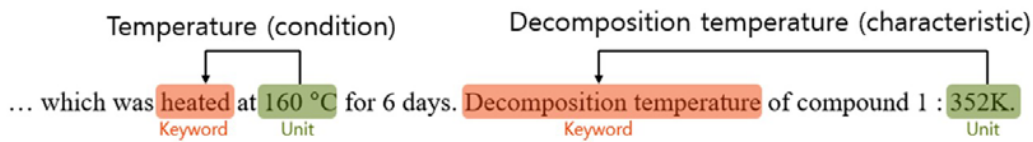

2) **Property classification method** :
   Using keyword, units are classified to each property.

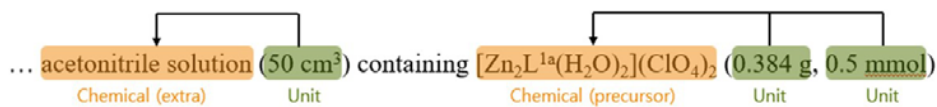

3) **Chemical matching method** :
   Distance based matching : Match with nearest Chemical or Operation

**Supplementary Figure S9**. Examples of condition extraction were obtained using an in-house Python-based rule-based code. (1) unit detection method, (2) property classification method, (3) chemical matching method.

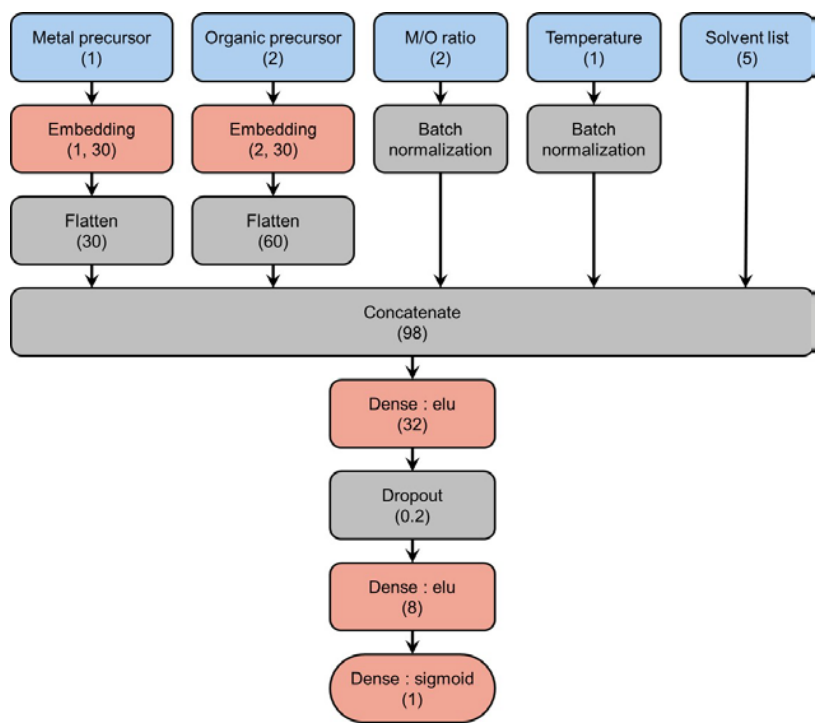

**Supplementary Figure S10**. A schematic diagram of the binary classifier of the PU learning algorithm.

**Supplementary Table S1**. Accuracy table of machine learning models for synthesis paragraph classification.

| Binary Classification | Precision | Recall | F1-score |
|---|---|---|---|
| logistic regression | 0.98 | 0.90 | 0.94 |
| soft vector machine | 1.00 | 0.85 | 0.92 |
| random forest | 1.0 | 0.67 | 0.80 |

**Supplementary Table S2**. Accuracy table of named entity recognition (NER) for chemicals.

| Chemical | Precision | Recall | F1-score |
|---|---|---|---|
| MOF | 0.89 | 0.93 | 0.91 |
| Precursor | 0.93 | 0.92 | 0.92 |
| Solvent | 0.91 | 0.90 | 0.90 |

**Supplementary Table S3**. The number of downloaded papers, format, and parser for each journal in article retrieval step.

| Journal | Number of papers | Format | Parser |
|---|---|---|---|
| American Chemical Society | 7,762 | XML | Self-developed parser |
| Elsevier | 9,467 | XML | Self-developed parser |
| Royal Society of Chemical | 5,000 | HTML | Self-developed parser |
| Springer | 1,691 | XML | Self-developed parser |
| Wiley | 4,645 | PDF | Chem data extractor parser |

**Supplementary Table S4**. Operations and its keywords extracted from synthetic paragraphs. Keywords are recognized using regular expressions. (..) means that it can exist in various forms after a word.

| Operation | Keyword | Example |
| --- | --- | --- |
| Heat | heat(..), oven, autoclave, Teflon-lined, solvothermal, hydrothermal | heating, heated |
| Cool | cool(..) | cooling, cooled |
| Stir | stir(..) | stirred, stirring |
| Wash | wash(..) | washed, washing |
| Remove | remov(..) | removing, removed |
| Dehydrate | dehydrat(..) | dehydrated |
| Desiccate | desiccat(..) | desiccated, desiccating |
| Dissolve | dissolve(..), redissolv(..) | dissolved, redissolved |
| Sonicate | sonic(..), ultrasonic(..) | sonicated, untrasonication |
| Diffuse | diffus(..) | diffused, diffusing |
| Store | stor(..) | storing, stored |
| Wait | wait(..), keep(..), kept, left | waited, keeping |
| Purify | purif(..) | purify, purification |
| Rinse | rins(..) | rinse, rinsed, rinsing |
| Filter | filter(..) | filtered, filtering |
| Dry | dri(..), dry(..) | dried, drying, dry |
| Ground | ground(..) | ground, grounding |
| Evaporate | evaporat(..) | evaporated, evaporating |
| Crystallize | crystalliz(..), recrystallize(..) | crystallized, recrystallizing |